\documentclass[prl,twocolumn,superscriptaddress]{revtex4-2}
\usepackage{graphicx}
\usepackage{bm, mathbbol}
\usepackage{hyperref}
\usepackage{amsmath,amssymb,amsfonts}
\usepackage{braket}
\usepackage{xcolor}
\usepackage[normalem]{ulem}

\begin{document}

\title{Interaction-driven dynamical quantum phase transitions in a strongly correlated bosonic system}

\author{Sebastian Stumper}
\email{sebastian.stumper@physik.uni-freiburg.de}
\affiliation{Institute of Physics, University of Freiburg, Hermann-Herder-Stra\ss e 3, 79104 Freiburg, Germany}

\author{Michael Thoss}%
\affiliation{Institute of Physics, University of Freiburg, Hermann-Herder-Stra\ss e 3, 79104 Freiburg, Germany}
\affiliation{EUCOR Centre for Quantum Science and Quantum Computing, University of Freiburg, Hermann-Herder-Stra\ss e 3, 79104 Freiburg, Germany}

\author{Junichi Okamoto}%
\affiliation{Institute of Physics, University of Freiburg, Hermann-Herder-Stra\ss e 3, 79104 Freiburg, Germany}
\affiliation{EUCOR Centre for Quantum Science and Quantum Computing, University of Freiburg, Hermann-Herder-Stra\ss e 3, 79104 Freiburg, Germany}

\date{\today}

\begin{abstract}

We study dynamical quantum phase transitions (DQPTs) in the extended Bose-Hubbard model after a sudden quench of the nearest-neighbor interaction strength. Using the time-dependent density matrix renormalization group, we demonstrate that interaction-driven DQPTs can appear after quenches between two topologically trivial insulating phases---a phenomenon that has so far only been studied between gapped and gapless phases. These DQPTs occur when the interaction strength crosses a certain threshold value that does not coincide with the equilibrium phase boundaries, which is in contrast to quenches that involve a change of topology. In order to elucidate the nonequilibrium excitations during the time evolution, we define a new set of string and parity order parameters. We find a close connection between DQPTs and these newly defined order parameters for both types of quenches. In the interaction-driven case, the order parameter exhibits a singularity at the time of the DQPT only when the quench parameter is close to the threshold value. Finally, the timescales of DQPTs are scrutinized and different kinds of power laws are revealed for the topological and interaction-driven cases.

\end{abstract}

\maketitle

\paragraph*{Introduction.}
The nonequilibrium dynamics of quantum many-body systems is a vibrant field of research, much less well understood than the phases of matter in thermal equilibrium. One way of bridging this gap is to drive a system away from an initial equilibrium state, for instance, by a sudden quench~\cite{mitraQuantumQuenchDynamics2016,esslerQuenchDynamicsRelaxation2016,langenPrethermalizationUniversalDynamics2016,buchholdPrethermalizationThermalizationQuenched2016}. This allows one to define new concepts of nonequilibrium criticality, and to study how these are related to the equilibrium cases~\cite{yuzbashyanRelaxationPersistentOscillations2006,barmettlerRelaxationAntiferromagneticOrder2009,ecksteinThermalizationInteractionQuench2009,sciollaQuantumQuenchesOffEquilibrium2010,mitraTimeEvolutionDynamical2012,heylDynamicalQuantumPhase2013,budichDynamicalTopologicalOrder2016,titumProbingGroundStatePhase2019}. Experimentally, cold atom systems can routinely implement a quantum quench, where a closed quantum system prepared in its ground state shows a nontrivial unitary evolution after a sudden change of parameters.~\cite{blochManybodyPhysicsUltracold2008,blochQuantumSimulationsUltracold2012,blattQuantumSimulationsTrapped2012,georgescuQuantumSimulation2014}.

In this context, a particularly interesting generalization of equilibrium concepts is that of dynamical quantum phase transitions (DQPTs), where time and the Loschmidt echo play the roles of temperature and the partition function, respectively~\cite{heylDynamicalQuantumPhase2013,heylDynamicalQuantumPhase2018}. As the Loschmidt echo is often inaccessible experimentally, it is highly desirable, from a practical viewpoint, to identify observables whose dynamics are linked to those of the Loschmidt echo.

It has been shown that, in general, there is no exact correspondence between DQPTs and the equilibrium quantum phase transitions of a system~\cite{andraschkoDynamicalQuantumPhase2014,vajnaDisentanglingDynamicalPhase2014,schmittDynamicalQuantumPhase2015,sharmaQuenchesDynamicalPhase2015}. On the other hand, various conditions have been identified under which a clear connection exists~\cite{heylDynamicalQuantumPhase2018}. Rigorous examples are exactly solvable models that exhibit ground states with distinct topological invariants~\cite{vajnaTopologicalClassificationDynamical2015a,huangDynamicalQuantumPhase2016,sedlmayrDynamicalPhaseTransitions2019}. Here a quench across the phase boundaries always leads to DQPTs. Moreover, after starting in symmetry broken phases, the respective order parameter becomes zero at the times of the DQPTs~\cite{heylDynamicalQuantumPhase2014,weidingerDynamicalQuantumPhase2017a,huangDynamicalQuantumPhase2019}. Both phenomena have been confirmed in experiments~\cite{jurcevicDirectObservationDynamical2017,zhangObservationManybodyDynamical2017,guoObservationDynamicalQuantum2019,flaschnerObservationDynamicalVortices2018,xuProbingDynamicalPhase2020}.

In contrast, the link of DQPTs with order parameter dynamics and equilibrium phase transitions is much more ambiguous for nonintegrable models~\cite{fogartyDynamicalPhaseTransitions2017,homrighausenAnomalousDynamicalPhase2017,halimehDynamicalPhaseDiagram2017,zunkovicDynamicalQuantumPhase2018}. Only recently, such correspondences were found for quenches in a nonintegrable spin-$1$ XXZ chain, with respect to the symmetry protected topological Haldane phase and its nonlocal string order parameter~\cite{hagymasiDynamicalTopologicalQuantum2019,peottaDeterminationDynamicalQuantum2020}. Such a model has also been experimentally realized~\cite{hilkerRevealingHiddenAntiferromagnetic2017,sompetRealisingSymmetryProtectedHaldane2021}. Further theoretical studies highlight the essential role played by symmetries in the dynamics of the string order parameter~\cite{mazzaOutofequilibriumDynamicsThermalization2014,calvanesestrinatiDestructionStringOrder2016}. It is yet unclear if these correspondences still hold when the symmetry is lowered or the quench does not cross a topological phase boundary.

In order to investigate such a problem, in this work, we study a chain of bosons with on-site and nearest-neighbor interactions~\cite{dallatorreHiddenOrder1D2006,bergRiseFallHidden2008, xuRealizingHaldanePhase2018}. Compared to the spin-$1$ XXZ chains, the model is less symmetric; for instance, particle-hole symmetry is missing. First, we consider quenches between topologically trivial phases. We find that strong nearest-neighbor interaction quenches above a threshold $V_c^\text{dyn}$ can induce DQPTs even when no topology is involved. We call these ``interaction-driven" DQPTs. Near the threshold, the order parameter develops nonanalytic signatures with temporal correspondence to the DQPTs. Second, we investigate quenches across topologically distinct phases starting from a Haldane insulator state. Here, quenches that reduce the interaction strength show no DQPTs on short timescales and no zeros of the order parameter, whereas quenches to stronger interactions do. The former, additionally, lead to a transient reduction of the entanglement entropy. Finally, we show that the time of the first DQPTs depends on the quench parameter roughly in a power law fashion for both kinds of initial states, while the detailed forms differ between the interaction-driven and topological cases.

\paragraph*{Model and methods.}
We consider the one-dimensional extended Bose-Hubbard model (EBHM) with on-site and nearest-neighbor interaction 
\begin{multline}
	\hat{H}= -J \sum_{i=1}^{L-1} \left(\hat{a}_i^\dagger \hat{a}_{i+1} + \text{H.c.} \right)  \\
	+ \frac{U}{2} \sum_{i=1}^{L} \hat{n}_i(\hat{n}_i - 1) + V \sum_{i=1}^{L-1} \hat{n}_i \hat{n}_{i+1} , \label{eq:HEBHM}
\end{multline}
where $\hat{a}_i^\dagger$ and $\hat{a}_i$ are bosonic creation and annihilation operators for site $i$ and $\hat{n}_i$ denotes the corresponding number operator. Throughout the paper, We fix the filling factor to $N/L = 1$ and the hopping strength to $J=1$, setting energy and time units.

This model has attracted considerable attention over the past decades, resulting in a good understanding of its equilibrium phases~\cite{dallatorreHiddenOrder1D2006,bergRiseFallHidden2008,dengEntanglementSpectrumOnedimensional2011,rossiniPhaseDiagramExtended2012,ejimaSpectralEntanglementProperties2014,gremaudExcitationDynamicsExtended2016}. These include a superfluid phase for weak interactions, a Mott insulator (MI) for dominating $U$, a density wave (DW) phase for dominating $V$, and the topologically nontrivial Haldane insulator (HI), which occurs for intermediate parameters. We focus on the case $U=5$, where the MI-HI (HI-DW) transition lies at $V^{\text{eq}}_{c1}\approx 2.95$ ($V^{\text{eq}}_{c2}\approx 3.525$)~\cite{rossiniPhaseDiagramExtended2012}.

We initialize the system in ground states corresponding to the MI ($V_\text{i}=1.0$) and HI ($V_\text{i}=3.25$) phases and subsequently drive it to nonequilibrium by a sudden change of $V$ from $V_\text{i}$ to $V_\text{f}$. The ground states are calculated by the density matrix renormalization group method, and the time-dependent variational principle is employed for the time evolution~\cite{haegemanTimeDependentVariationalPrinciple2011,haegemanUnifyingTimeEvolution2016,jaschkeOpenSourceMatrix2018a}. We have checked that the results are independent of the boundary conditions~\cite{stumperMacroscopicBoundaryEffects2020}. Also we note that the initial states are neither simple Fock states nor highly symmetric states such as at the Affleck-Kennedy-Lieb-Tasaki (AKLT) or Heisenberg points in the Haldane phase of spin-1 chains~\cite{mazzaOutofequilibriumDynamicsThermalization2014,calvanesestrinatiDestructionStringOrder2016,hagymasiDynamicalTopologicalQuantum2019,wozniakRelaxationDynamicsSpin12016}. As shown below, the initial condition sensibly affects the DQPTs.

\paragraph*{Quantities of interest.}
One way to generalize the notion of thermodynamic phase transitions to the nonequilibrium case is to consider the rate function
\begin{equation}
	\lambda_L(t) = -\frac{1}{L} \ln |\braket{\psi(0)|\psi(t)}|^2
\end{equation}
as a dynamical analog of the free energy density, where $\ket{\psi(t)}=e^{-\text{i}\hat{H}t}\ket{\psi(0)}$  and $\ket{\psi(0)}$ is the initial wavefunction of the system. DQPTs can be defined as nonanalytic points of this function in the thermodynamic limit~\cite{heylDynamicalQuantumPhase2013}. 

Second, we study the time evolution of the order parameters to quantify the relation of the time-evolved states to the underlying ground state orders. For the MI and HI phases, these are, respectively, described by the nonlocal string and parity operators~\cite{dallatorreHiddenOrder1D2006}
\begin{equation}
	\begin{split}
		\hat{O}^z_\mathrm{string}(i,j) &= \delta \hat{n}_i \Big( \prod_{i<k<j} e^{i \pi \delta \hat{n}_k} \Big) \delta \hat{n}_j, \\
		\hat{O}^z_\mathrm{parity}(i,j) &= \Big( \prod_{i<k<j} e^{i \pi \delta \hat{n}_k} \Big),
	\end{split}
	\label{eq:Ostr}
\end{equation}
where $\delta \hat{n}_i = \hat{n}_i - 1$. $\hat{O}^z_\text{string}$ is inspired by the string operators corresponding to the $z$ component of spin-$1$ XXZ chains, which can be considered as an effective model of the EBHM under the mapping  $\delta n_i \rightarrow S_i^z = -1,0,1$. Order parameters are defined as the long-distance limits, i.e., $O_\gamma =  \lim_{|i-j|\to\infty} \braket{\hat{O}_\gamma (i,j)}$ for $\gamma\in\{\text{string, parity}\}$. Below, we use $i = L/4$ and $j=3L/4$.

We further propose another set of nonlocal operators that correspond to the $x$ components of the spin-1 operator as
\begin{equation}
	\begin{split}
		\hat{O}_\text{string}^x(i,j) &= \frac{1}{4} \left(\hat{a}_i + \hat{a}_i^\dagger \right) \Big(\prod_{i<k<j} \hat{O}_k \Big) \left(\hat{a}_j + \hat{a}_j^\dagger \right) ,\\	
		\hat{O}_\text{parity}^x(i,j) &= \Big(\prod_{i<k<j} \hat{O}_k \Big),
	\end{split}
	\label{eq:Ox}
\end{equation}
where
\begin{equation}
	\hat{O}_k = (-1) \cdot P_k^{\leq 2} e^{i\pi S_k^x} P_k^{\leq 2}, \qquad 
	e^{i\pi S_k^x}
	=
	\begin{pmatrix}
		0 & 0 & -1 \\
		0 & -1 & 0 \\
		-1 & 0 & 0
	\end{pmatrix},
\end{equation}
and $P_k^{\leq 2}$ is the projection onto site occupations $n_k \leq 2$. \footnote{We have to introduce a minus sign in $\hat{O}_k$ compared to the spin-$1$ model, because the signs of the bosonic hopping term $\sim (-J)$ and the spin-$1$ XX term are opposites~\cite{pollmannEntanglementSpectrumTopological2010}. See also supplemental material for a discussion.} In principle, the eigenvalues of $\hat{O}_k$ should all be of modulus $1$, because otherwise the long-distance limit of expectation values of Eq.~\eqref{eq:Ox} will be either $0$ or $\infty$~\cite{perez-garciaStringOrderSymmetries2008}. However, the above definition always yields a zero order parameter in the presence of occupation numbers $n>2$. This problem cannot be mitigated by a straightforward generalization of $\hat{O}_k$ to $n>2$, because there are no negative counterparts $2-n<0$. More details are given in the supplemental material (SM) \footnote{See supplemental material at [URL will be inserted by publisher]}. 

A simple solution consists in using the following projection string operator
\begin{equation}
	\hat{P}(i,j) = \Big( \prod_{i<k<j} \hat{P}_k^{\leq 2} \Big).
\end{equation}
As shown in SM~\cite{Note2}, $\hat{P}(i,j)$ exhibits the same asymptotic decay length $\xi$ as the $x$ component operators of Eq.~\eqref{eq:Ox}, i.e.,
\begin{equation}
	\hat{P}(i,j) \sim \hat{O}_\gamma^x(i,j) \sim e^{ -\xi |i-j| }
\end{equation}
for $\gamma\in\{\text{string, parity}\}$. Hence, we obtain well-defined order parameters
\begin{equation}
	\tilde{O}_\gamma^x = \lim_{|i-j|\to\infty} \frac{\braket{\hat{O}_\gamma^x(i,j)}}{\braket{\hat{P}(i,j)}}.
\end{equation}
As the operators in Eq.~\eqref{eq:Ox} are off-diagonal in the Fock basis, these order parameters represent a new type of long-range phase coherence.

\begin{figure}[t]
	\centering
	\includegraphics[width=\columnwidth]{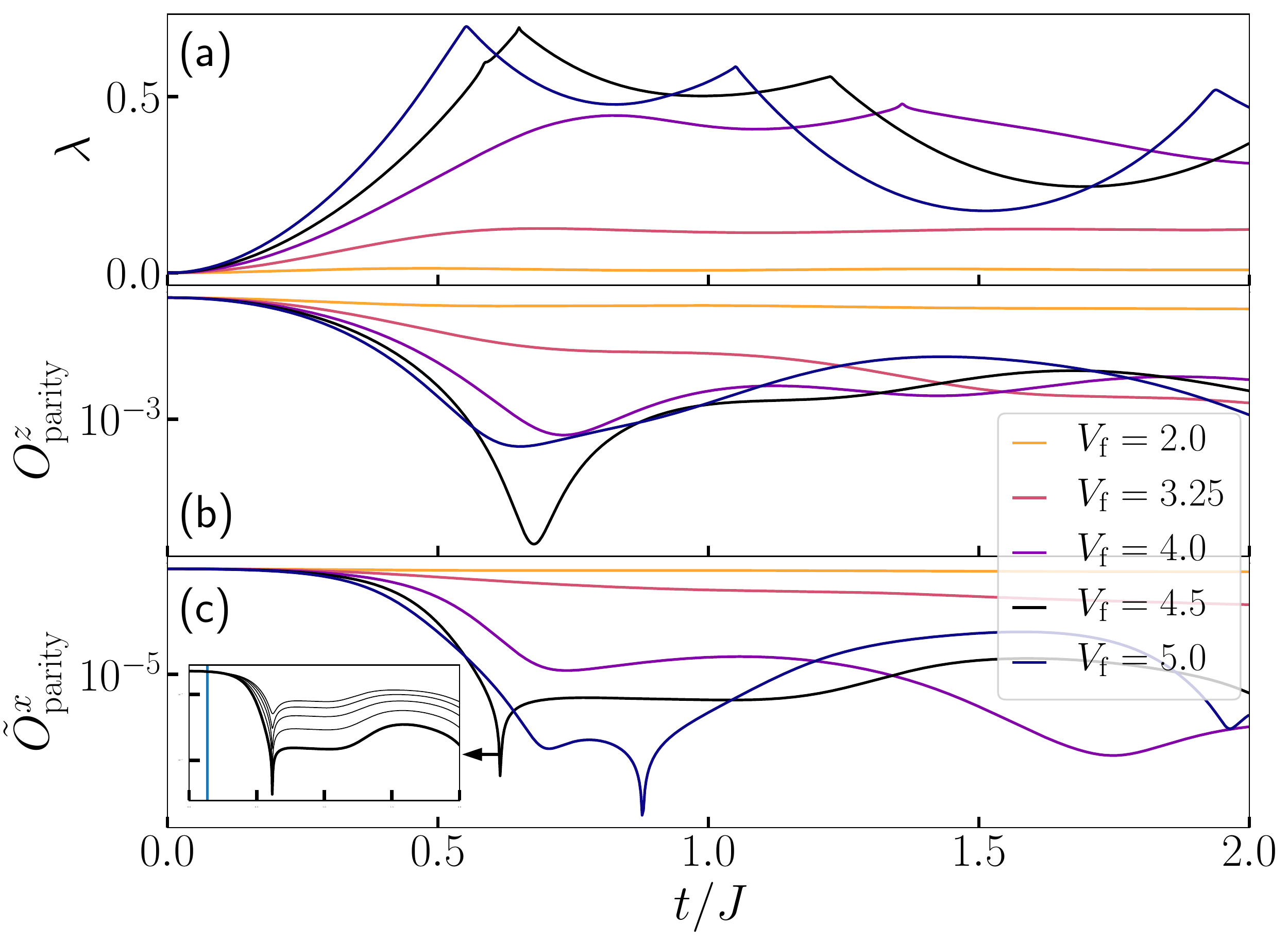}
	\caption{Time evolution of (a) the rate function, (b) the $z$ parity order, and (c) the renormalized $x$ parity order. The initial state is in the MI phase ($V_\text{i}=1.0$), and  quenched to $V_\text{f}$ in the MI ($2.0$), HI ($3.25$) and CDW ($4.0,4.5,5.0,6.0$) phases. $L=270$ is used. The inset in panel (c) shows the case $V_\text{f}=4.5$ for system sizes $L=54,80,120,180,270$. The axes are the same as for panel (c).}
	\label{fig:MI}
\end{figure}

\paragraph*{Results.} 
We now turn to the numerical results. Fig.~\ref{fig:MI} shows the time evolution of the system initialized in an MI ground state ($V_\text{i} = 1.0$) and quenched to $V$. As depicted in Fig.~\ref{fig:MI}(a), the rate function develops pronounced kinks when $V_\text{f}$ is above 
certain threshold values. For the first maximum of the rate function we find  $V_c^\text{dyn} \approx 4.2$ (see Fig.~\ref{fig:tstar}), for the second one $V_c^\text{dyn} \approx 3.6$. Both cases represent interaction-driven DQPTs, as the critical interaction strengths to observe DQPTs are not related to the topological MI-HI transition at $V^{\text{eq}}_{c1} \approx 2.95$. The former, $V_c^\text{dyn} \approx 4.2$, is also far 
away from the HI-CDW phase boundary at $V^{\text{eq}}_{c2} \approx 3.525$.

The corresponding dynamics of $O_\text{parity}^z$ and $\tilde{O}_\text{parity}^x$ are shown in  Figs.~\ref{fig:MI}(b)-(c), respectively. The two parity order parameters remain close to their initial values when $V_\text{f}$ is in the MI region ($V_\text{f} = 2.0$). On the other hand, they decay by several orders of magnitude, if the phase boundary is crossed, i.e., $V_\text{f} \geq V_{c1}^\text{eq} \approx 2.95$. In particular, when $V_\text{f}$ is just above the threshold value of the DQPTs, $V_c^\text{dyn} \approx 4.2$, the parity order parameters show minima at the time of the first kink of the rate function. For $\tilde{O}_\text{parity}^x$ the minimum is especially sharp. After the second kink, when $\lambda$ drops to lower values again, the order parameters show revivals. For larger $V_\text{f}$, there is no longer such a clear temporal relation (see SM~\cite{Note2} for a finer grid around $V_\text{f}=4.5$). As shown in SM~\cite{Note2}, we confirmed similar behavior of the parity order parameters in a spin-1 chain. Contrarily, for the DQPTs for a quench starting from the topologically nontrivial HI phase (shown below), the temporal correspondence between the rate function and the order parameters persists for any quench, not just near $V_c^\text{dyn}$.

Interestingly, a finite-size analysis reveals that $\tilde{O}_\text{parity}^x$ actually vanishes for all $V_\text{f}$ after a certain time as shown in the inset of Fig.~\ref{fig:MI}(c), where the time is marked by a vertical blue line. This means that the correlation function $\langle \hat{O}_\text{parity}^x(i,j) \rangle$ decays more rapidly with $|i-j|$ than the projection string $\langle \hat{P}(i,j) \rangle$, and the quench destroys the corresponding long-range phase coherence. The time evolution shown in Fig.~\ref{fig:MI}(c), therefore, indicates only short-range correlations. By contrast, in the spin-$1$ model, $O_\text{parity}^x$ does not vanish in the thermodynamic limit~\cite{Note2}.

\begin{figure}[t]
	\centering
	\includegraphics[width=\columnwidth]{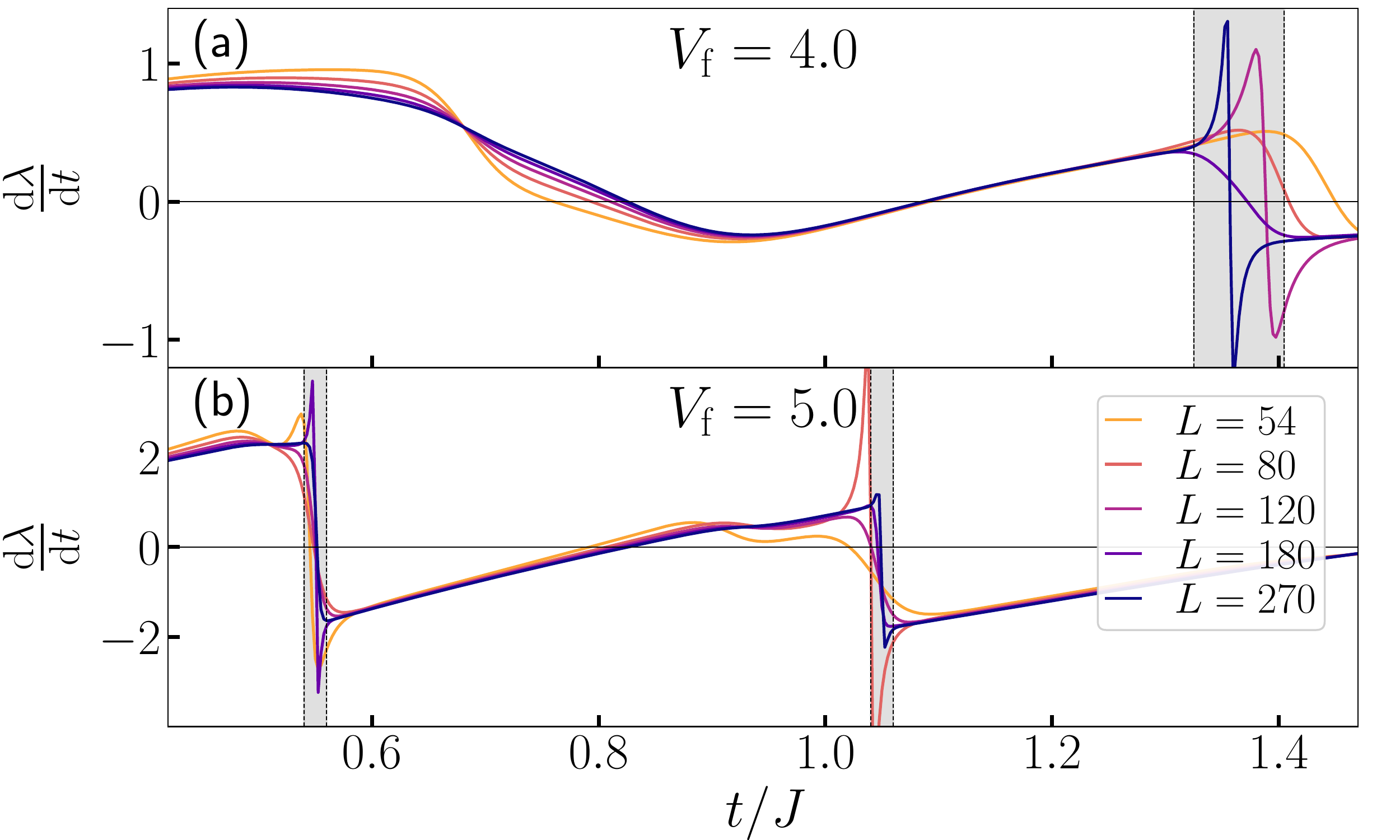}
	\caption{Finite-size scaling of $\mathrm{d}\lambda/\mathrm{d}t$ for (a) $V_\text{f}=4.0$ and (b) $V_\text{f}=5.0$, with the initial state in the MI phase ($V_\text{i}=1.0$). Shaded regions indicate where the rate function depends sensitively on $L$.}
	\label{fig:MI_FSS}
\end{figure}

In order to confirm that $\lambda$ develops nonanalytic features at certain critical times $t^*$, we analyze the system-size dependence of $\mathrm{d}\lambda/\mathrm{d}t$ in Fig.\ref{fig:MI_FSS}. For $V_\text{f}=4.0$ [Fig.~\ref{fig:MI_FSS}(a)], the first maximum around $t\approx 0.8$ is smooth, while the second one converges slowly in $L$ towards a sharp jump. For the stronger quench to $V_\text{f}=5.0$ [Fig.~\ref{fig:MI_FSS}(b)], $\mathrm{d}\lambda/\mathrm{d}t$ changes signs at critical times $t^* / J \approx 0.55$ and $1.05$. With increasing $L$, the regions around the jumps showing strong system-size dependencies diminish. For example, for $V_\text{f}=5.0$ at $t^* / J \approx 0.55$, there are large spikes in $\mathrm{d}\lambda/\mathrm{d}t$ only for $L=54$ and $180$.

\begin{figure}[t]
	\centering
	\includegraphics[width=\columnwidth]{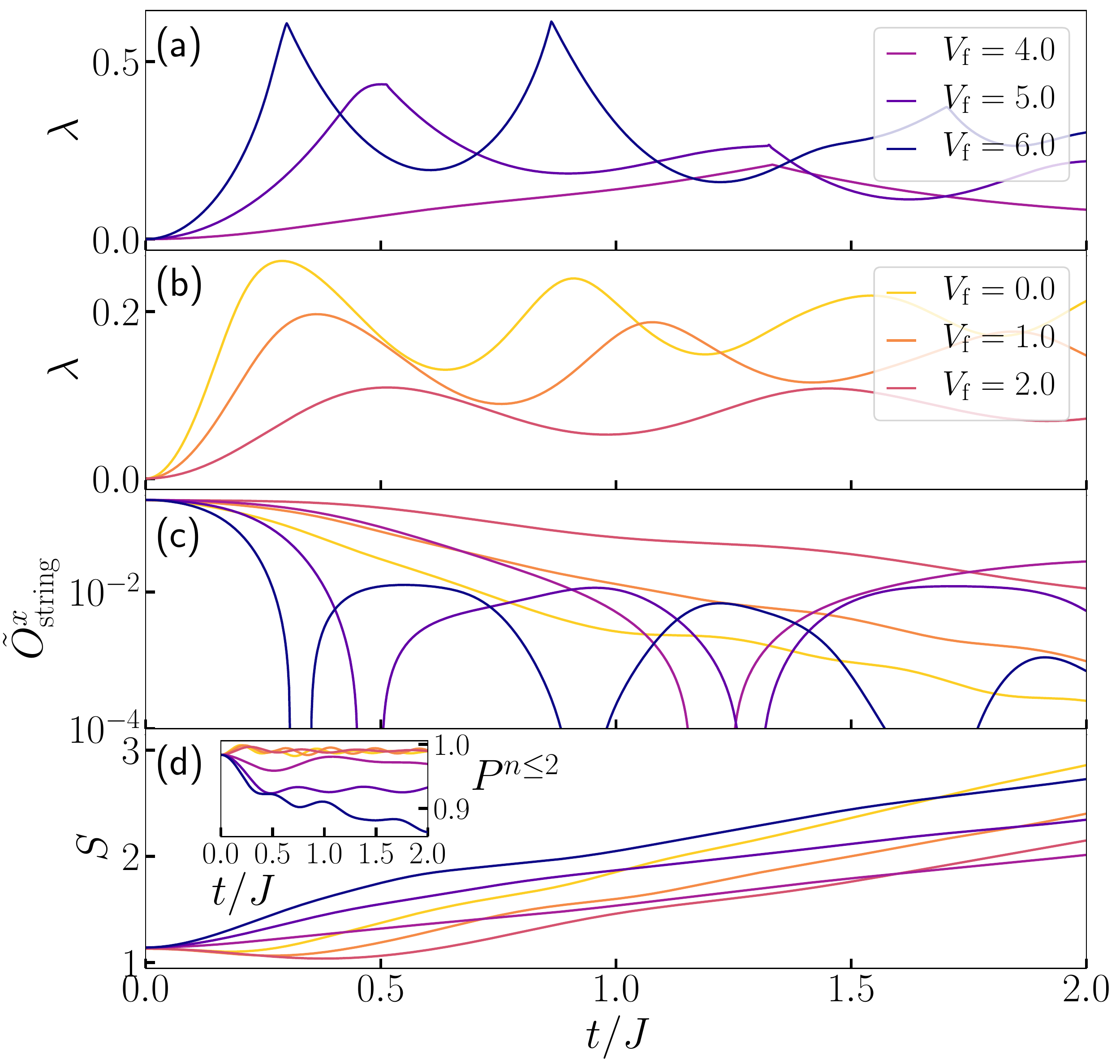}
	\caption{Time evolution of (a), (b) the rate function for quenches to larger and lower $V_\text{f}$'s, respectively, (c) the $x$ string order, and (d) the entanglement entropy. The initial state is in the HI phase ($V_\text{i}=3.25$), and quenched to $V_\text{f}$ in the MI ($0.0,1.0,2.0$) and CDW ($4.0,5.0,6.0$) phases. $L=600$ is used for $\lambda$, for all other observables $L=270$. The inset of panel (d) shows the projection to $n\leq 2$ per site.}
	\label{fig:HI}
\end{figure}

Next, let us turn to the case where the system is initialized in the HI phase (Fig.~\ref{fig:HI}). We always find DQPTs when $V$ is quenched to a larger value across the HI-CDW boundary, as shown in Fig.~\ref{fig:HI}(a). Thus, we have $V_c^\text{dyn}=V_{c2}^\text{eq}$. On the other hand, for quenches across the MI-HI boundary to $V_\text{f}<V_{c1}^\text{eq}$, depicted in Fig.~\ref{fig:HI}(b), there are no kinks on the given timescale. While it is possible that DQPTs occur at later times, we certainly see that the first few maxima of $\lambda$ are smooth. This result is in contrast to Ref.~\cite{hagymasiDynamicalTopologicalQuantum2019}, which always found kinks at the maxima of $\lambda$ for a spin-1 XXZ chain initialized to the AKLT state within the Haldane phase. The discrepancy can be traced back to the lower symmetry of our initial state. As shown in SM~\cite{Note2}, a spin-1 chain initialized in a less symmetrical state but still in the Haldane phase shows smooth first maxima of $\lambda$ for corresponding quenches.

The dynamics of the renormalized $x$ string order parameter $\tilde{O}_\text{string}^x$ are depicted in Fig.~\ref{fig:HI}(c). We find zeros of $\tilde{O}_\text{string}^x$ in close temporal relation to the kinks of the rate function. Similar results were obtained for the spin-1 chains~\cite{hagymasiDynamicalTopologicalQuantum2019}. In the presented time range, the zeros appear only when $V$ is increased, but not when $V$ is lowered. However, a longer time simulation up to $t/J=4.0$ for a quench from HI to MI shows a zero of $\tilde{O}_\text{string}^x$, indicating the existence of DQPTs~\cite{Note2}.

As another interesting property of the time-evolving state, we consider the entanglement entropy $S$ about the central bond, which we depict in Fig.~\ref{fig:HI}(d). While it shows the typical approximately linear increase at later times, the short-time behavior depends on whether we lower or increase the interaction $V$. The entanglement entropy increases monotonically for increased $V$ also at short times. A decreased $V$, on the other hand, leads to a decrease of entanglement in the system at short times. The weaker the quench, the stronger is the suppression of the entanglement. As shown in the inset of Fig.~\ref{fig:HI}(d), the projections to $n\leq 2$ increase when the entanglement entropy decreases, because of the lowered nearest-neighbor repulsion. The entanglement entropy of the ground state is lower in the MI phase along the parameters considered~\cite{dengEntanglementSpectrumOnedimensional2011}. Thus, we infer that at short times the suppression of higher occupations reduces the entanglement entropy. Other possible observables, such as the doublon density, are likely to behave similarly. The following increase of the entanglement entropy is due to a slower build up of long-range  correlations.

\begin{figure}[t]
	\centering
	\includegraphics[width=\columnwidth]{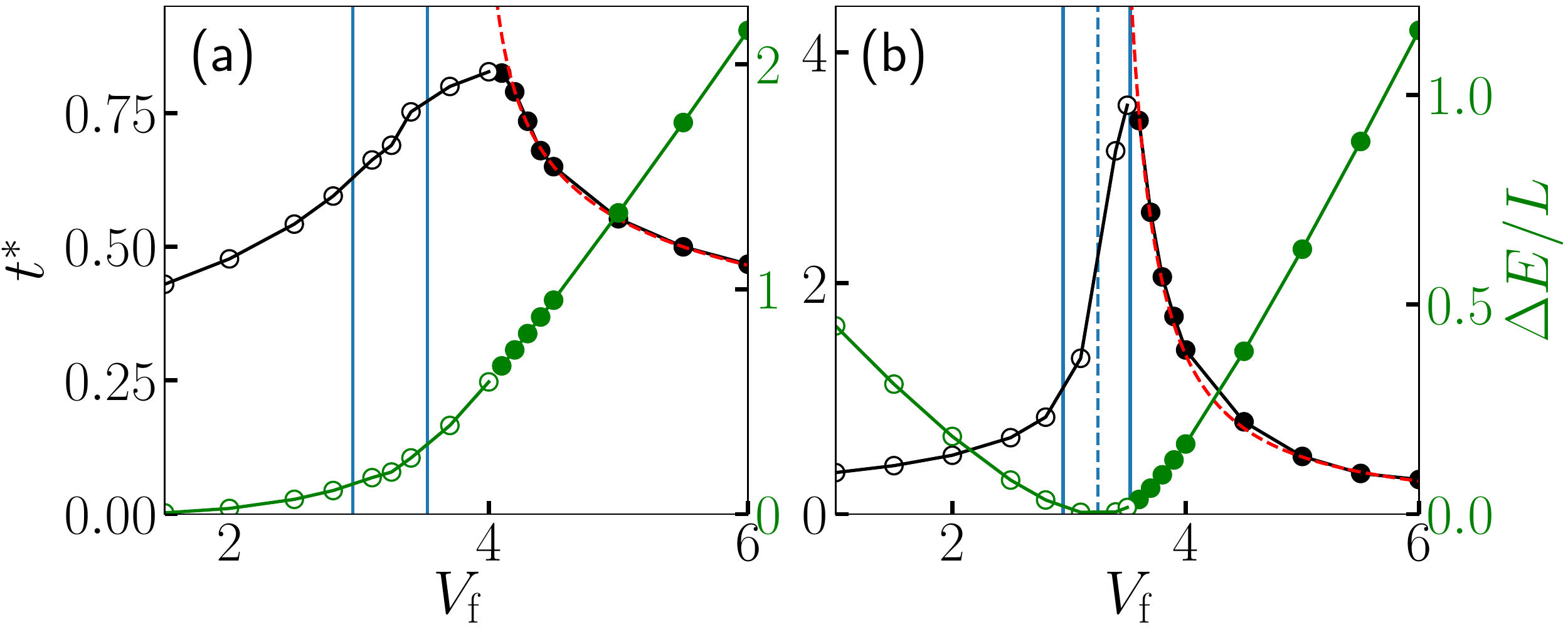}
	\caption{Time $t^*$ of the first maxima of $\lambda$ (left axes, black symbols) and energy density relative to the ground state of the quenched Hamiltonian (right axes, green symbols) are plotted against $V_f$. Full circles correspond to kinks, and empty circles to smooth maxima. Panels (a) and (b) correspond to initial states in the MI or HI phase (see Figs.~\ref{fig:MI} and \ref{fig:HI}, respectively. Solid vertical lines indicate the MI-HI and HI-CDW transitions, while the dashed line in (b) shows $V_\text{i}$. The red dashed lines are power law fits, (a) $t^*\approx 0.56 \cdot (V_\text{f}-3.92)^{-0.26}$ and (b) $t^*\approx (V_\text{f}-V_\text{i})^{-1.2}$.}
	\label{fig:tstar}
\end{figure}

Finally, we discuss time $t^*$ of the first maxima in $\lambda$ and quench-induced heating $\Delta E / L$ with respect to the quenched interaction (see Fig.~\ref{fig:tstar}). We have checked that the latter is converged in the system size despite the open boundary condition. Here, $\Delta E = \braket{\psi(t) | H_\text{f} | \psi(t)} - E_0^\text{f}$ is the energy of the time-evolving state relative to the ground state of the quenched Hamiltonian $H_\text{f}$. Since the system is isolated, the energy is constant for $t>0$.

The induced energy densities (green) behave approximately linearly for $V_\text{f}\gtrsim V^\text{eq}_{c2}$ for initial states in the MI  [Fig.~\ref{fig:tstar}(a)] and HI  [Fig.~\ref{fig:tstar}(b)] phases. The values and the slope are larger for the initial state in the MI phase. This larger slope is a consequence of a higher doublon density in the HI phase, which reduces the extra energy due to a given increase of $V_\text{f}$.

Next, we look at the times $t^*$ of the first maximum of $\lambda$, which is either smooth (empty circles) or nonanalytic (full circles). For both kinds of initial states, $t^*$ attains a maximum at the threshold value $V_c^\text{dyn}$, where the nonanalyticities start to appear. As $V_\text{f}$ increases beyond $V_c^\text{dyn}$, the time $t^*$ decays in a power law manner, as indicated by the fits in Fig.~\ref{fig:tstar} (dashed red lines). Heuristically, such power laws can be understood in terms of the width of the spectra of $\ket{\psi_\text{i}}$ with respect to the eigenstates of $H_\text{f}$ (see SM~\cite{Note2}), which broaden as the quenches become stronger. Interestingly, starting from an HI state leads to an apparent divergence of the extrapolation of the power law at $V_\text{i}$, $t^*\sim (V_\text{f}-V_\text{i})^{-1.2}$. Such an approximate power law with a divergence at the initial parameter is typical of DQPTs in noninteracting topological systems~\cite{heylDynamicalQuantumPhase2013,karraschDynamicalPhaseTransitions2013,budichDynamicalTopologicalOrder2016,bhattacharjeeDynamicalQuantumPhase2018}. Thus, the present results further underpin the topological nature of quenches starting from the HI phase. On the other hand, for the interaction-driven DQPTs, which occur with initial states in the MI phase, $t^*$ is not well approximated by a simple power law in $(V_\text{f} - V_\text{i})$.

\paragraph*{\label{sec:con}Conclusions.}
In this work we have analyzed the dynamics of the extended Bose-Hubbard model after sudden interaction quenches. We have contrasted initial states in the Mott and Haldane insulator phases. In case of the Mott insulator, DQPTs are induced by sufficiently strong quenches. These DQPTs correspondence to the equilibrium phase boundaries and bear no relation to topology. Near the threshold quench value, non-analytic signatures of the parity order parameter accompany the DQPTs. Starting from the Haldane insulator, we find DQPTs and zeros of the string order parameter for quenches to larger nearest-neighbor interactions. While, for quenches to lower interactions, the short time behavior differs from a previous study~\cite{hagymasiDynamicalTopologicalQuantum2019}, we expect that DQPTs occur for longer times. The discrepancy is attributed to different symmetries of initial states. Finally, we have shown that the timescales of DQPTs depend on the quench parameter in a power law manner, and its precise form differs between two types of DQPTs--- either topological or interaction-driven. Experimental tests of these results by ultracold atoms could be feasible in the near future. On the theoretical side, further study of the timescales of the dynamics is an intriguing open problem.

\begin{acknowledgments}
\paragraph*{Acknowledgments.}
The authors would like to thank Markus Heyl for fruitful discussions and acknowledge support from the Georg H. Endress Foundation. Furthermore, support by the state of Baden-W\"urttemberg through bwHPC and the German Research Foundation (DFG) through Grant No. INST 40/467-1 FUGG is acknowledged. 
\end{acknowledgments}



%

\end{document}


\title{Supplemental material for ``interaction-driven dynamical quantum phase transitions in a one-dimensional boson system"}

\author{Sebastian Stumper}
\affiliation{Institute of Physics, University of Freiburg, Hermann-Herder-Stra\ss e 3, 79104 Freiburg, Germany}

\author{Michael Thoss}%
\affiliation{Institute of Physics, University of Freiburg, Hermann-Herder-Stra\ss e 3, 79104 Freiburg, Germany}
\affiliation{EUCOR Centre for Quantum Science and Quantum Computing, University of Freiburg, Hermann-Herder-Stra\ss e 3, 79104 Freiburg, Germany}

\author{Junichi Okamoto}%
\affiliation{Institute of Physics, University of Freiburg, Hermann-Herder-Stra\ss e 3, 79104 Freiburg, Germany}
\affiliation{EUCOR Centre for Quantum Science and Quantum Computing, University of Freiburg, Hermann-Herder-Stra\ss e 3, 79104 Freiburg, Germany}

\date{\today}

\maketitle

\section{Renormalized $x$-string and $x$-parity order parameters for the extended Bose-Hubbard model}
\begin{figure}[h]
	\centering
	
	\begin{minipage}{0.45\textwidth}
		\centering
		\includegraphics[width=\textwidth]{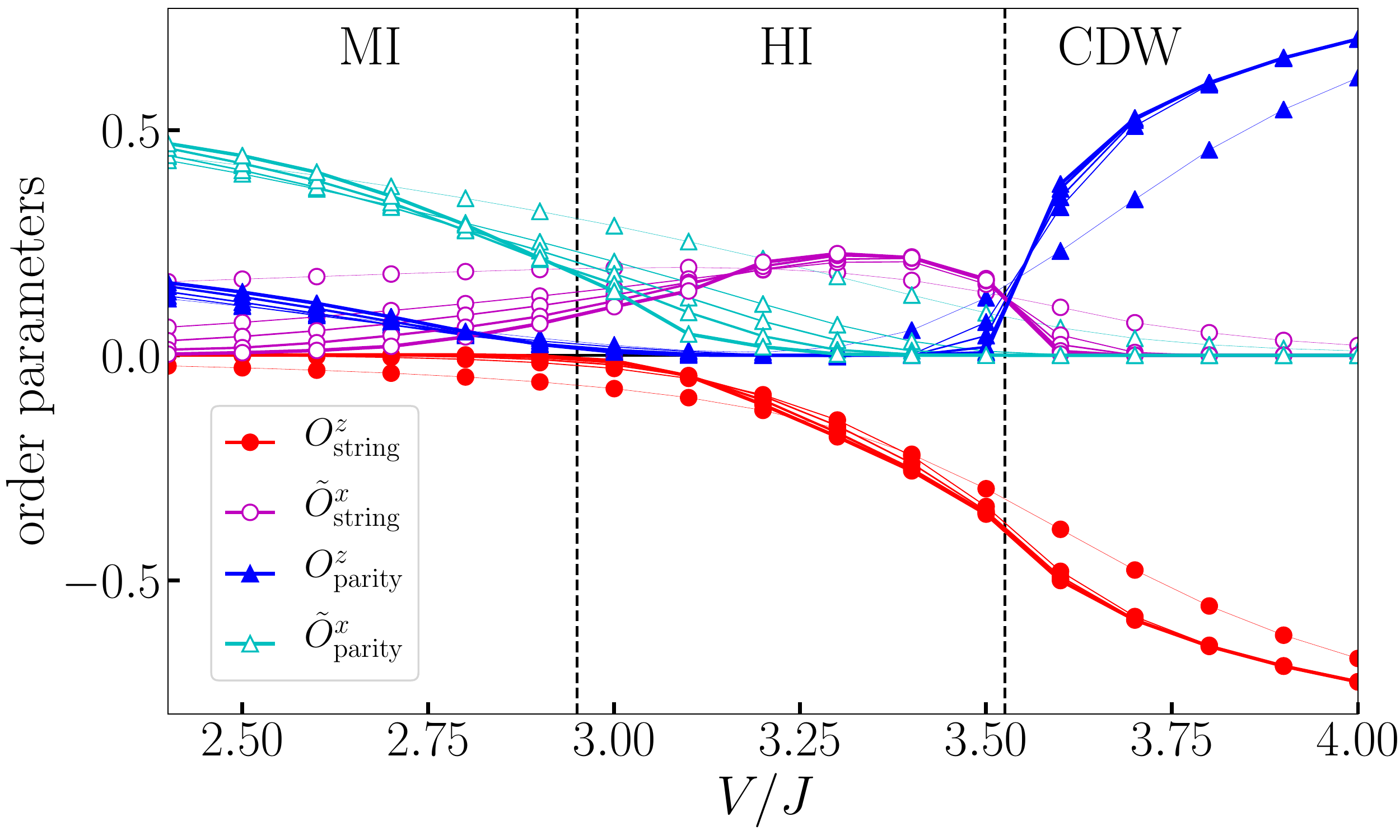}
		\caption{Phase diagram and the corresponding order parameters for the extended Bose Hubbard model. Usual $z$-component order parameters and renormalized $x$-component order parameters as a function of $V$. For each, we show system sizes $L=54,80,120,180$, increasing from thin to thick lines. The dashed vertical lines mark the phase boundaries according to Ref.~\cite{rossiniPhaseDiagramExtended2012}.}
		\label{fig:phase_OP}
	\end{minipage}
	\hfill
	\begin{minipage}{0.45\textwidth}
		\centering
		\includegraphics[width=\textwidth]{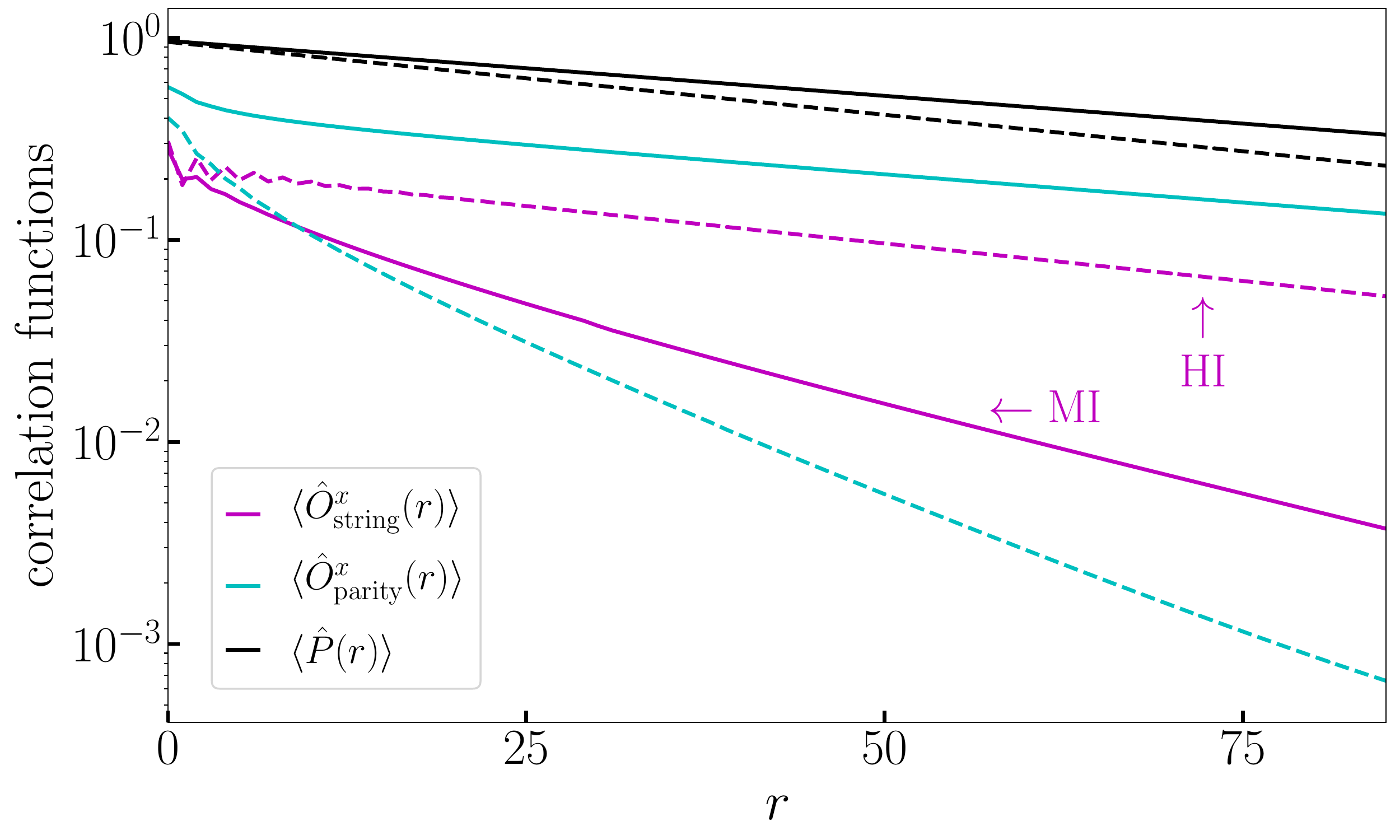}
		\caption{$x$-string and $x$-parity correlation functions without renormalization factor as a function of $r$, as well as the projection string $P(r)$. Dashed lines correspond to $V=2.5$ (MI) and solid lines to $V=3.2$ (HI). \vspace{3\baselineskip} }
		\label{fig:phase_Corr}
	\end{minipage}
\end{figure}

We show that the renormalized order parameters, which are introduced in Eqs.~(4)-(8) of the main text, provide a well defined way of determining the phase diagram. In Fig.~\ref{fig:phase_OP}, both the $x$- and $z$- components of the order parameters, $O_\text{string,parity}^z$ and $\tilde{O}_\text{string,parity}^x$, are depicted as a function of $V$ for increasing system sizes. Even though there is no SU(2) symmetry in the extended Bose-Hubbard model (EBHM) due to the lack of particle-hole symmetry, the order parameters of both components clearly distinguish the different phases. The Mott insulator (MI) is characterized by non-vanishing parity order parameters, while in the Haldane insulator (HI) phase, the string order parameters take finite values.

Contrary to the renormalized $x$-components, the bare $x$-components of the order parameters are insufficient to distinguish different phases. In Fig.~\ref{fig:phase_Corr}, we show the correlation functions $\langle \hat{O}_\text{string,parity}^x(r)\rangle$ and the projection string operator $\langle \hat{P}(r) \rangle$ over the distance $r=|i-j|$ between two sites for selected points in the MI (full lines) and HI (dashed lines) phase diagram. All correlation functions decay exponentially, as in Eq.~(7) of the main text. For the projection string $\hat{P}(r) = \prod_{i < k < i+r} P^{n_k\leq 2}$ the decay length is given by 
\begin{equation}
\xi = -\log \langle P^{n\leq 2} \rangle,
\end{equation}
i.e., for each site the projection to $n\leq 2$ contributes the same factor. In the MI phase, the decay length of $\langle \hat{O}_\text{parity}^x(r)\rangle$ is equal to $\xi$, whereas in the HI phase it is smaller than $\xi$. Therefore, in the MI phase, the decay of $\langle \hat{O}_\text{parity}^x(r)\rangle$ is exclusively due to the projections in Eq.~(5) of the main text, and we consider the system to have $x$-parity order in the renormalized sense. For $\langle \hat{O}_\text{string}^x(r)\rangle$, the decay length is given by the projection in the HI phase, while it decays faster due to its own fluctuations in the MI phase.

Naturally, it is desirable to have a definition of $x$-parity and $x$-string order that does not rely on projections. However, this is not trivial. Note that the $x$ operators are mainly products of 
\begin{equation}
	\hat{O_k} = \ket{2}\bra{0} + \ket{1}\bra{1} + \ket{0}\bra{2},
\end{equation}
acting on each site $k$, by analogy to $e^{i \pi \hat{S}_k^x}$ [see Eq.~(5) of the main text]. A straightforward generalization to $n>2$, would require terms like $\ket{2-n}\bra{n}$ with $2-n<0$. Another simple ansatz, adding the terms $\ket{n}\bra{n}$ for $n>2$, also fails. Here the reason is nonlocality. Because the filling factor is $\overline{n}=1$, every site with $n>2$ requires that there are more sites with $n=0$ than with $n=2$. The application of $\prod_k ( \hat{O}_k + \sum_{n_k>2} \ket{n_k}\bra{n_k})$ would then change the total particle number and therefore yield an expectation value of zero, if any site has $n>2$. 

In conclusion, on the one hand, such operators may exist that correspond to the $x$-string and $x$-parity operators of spin-$1$ chains without prescriptions of projections. On the other hand, it is unclear whether these operators can be expressed simply as products of local operators.

\section{Long-time dynamics for the HI-CDW quench with $V_\text{f} \simeq V_c^\text{dyn}$}
\begin{figure}[h]
	\centering
	\includegraphics[width=0.5\textwidth]{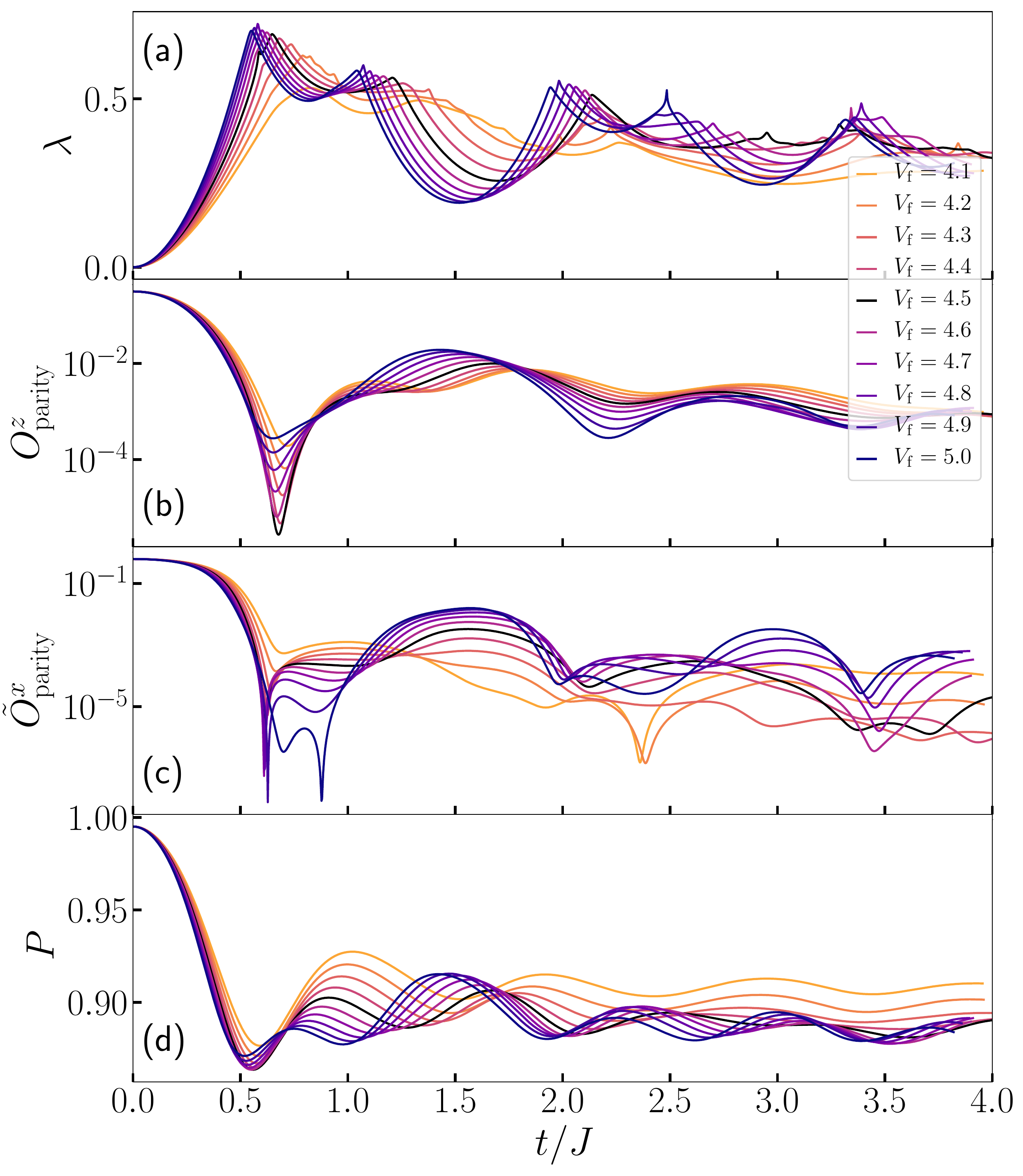}
	\caption{Dynamics of (a) the rate function, (b) and (c) the parity order parameters, and (d) the projection to $n=0,1,2$ per site. The initial state is in the MI phase ($V_\text{i}=1.0$), and various $V_\text{f}$ close to $V_c^\text{dyn}$ are used. The system size is $L=120$.}
	\label{fig:gridMI}
\end{figure}

Here, for the case of an initial state in the MI phase, we provide a finer parameter grid around the threshold value $V_\text{f}\simeq V_c^\text{dyn}$ for longer times up to $t/J=4$ with $L=120$ (see Fig.~\ref{fig:gridMI}). We clearly see the relation between the time evolution of the rate function [Fig.~\ref{fig:gridMI}(a)] and that of the order parameters [Fig.~\ref{fig:gridMI}(b)-(d)]. The nonlocal order parameters $O_\text{parity}^z$ and $\tilde{O}_\text{parity}^x$ develop a pronounced minimum for values of $V_\text{f}$ when the first maximum of $\lambda$ becomes nonanalytic, i.e., at a dynamical quantum phase transition (DQPT). Moreover, we see that $\lambda$ has pairs of DQPTs and downturns between two such pairs. These downturns of $\lambda$ are accompanied by revivals of the order parameters. On the other hand, for $V_\text{f}=5.0 > V_c^\text{dyn}$ shown in the main text, there is no close connection between the minimum of $\tilde{O}_\text{parity}^x$ and the first DQPT.

\section{Long-time dynamics for the HI-MI quench}
In order to show that DQPTs likely exist for HI to MI quenches, we show the results of a simulation up to $t/J=5$ with $V_\text{i}=3.25$ and $V_\text{f}=1.0$. While the finite size effects are too strong to reveal a kink of $\lambda$, there is clearly a zero of $\tilde{O}_\text{string}^x$ close to $t/J=4$. This indicates that DQPTs occur also for the HI-MI quench in long-time simulations. Furthermore, as in the inset of Fig.~2(c) of the main text, we see that the order parameter vanishes in the thermodynamic limit after a certain time.
\begin{figure}[h]
	\centering
	\includegraphics[width=0.5\linewidth]{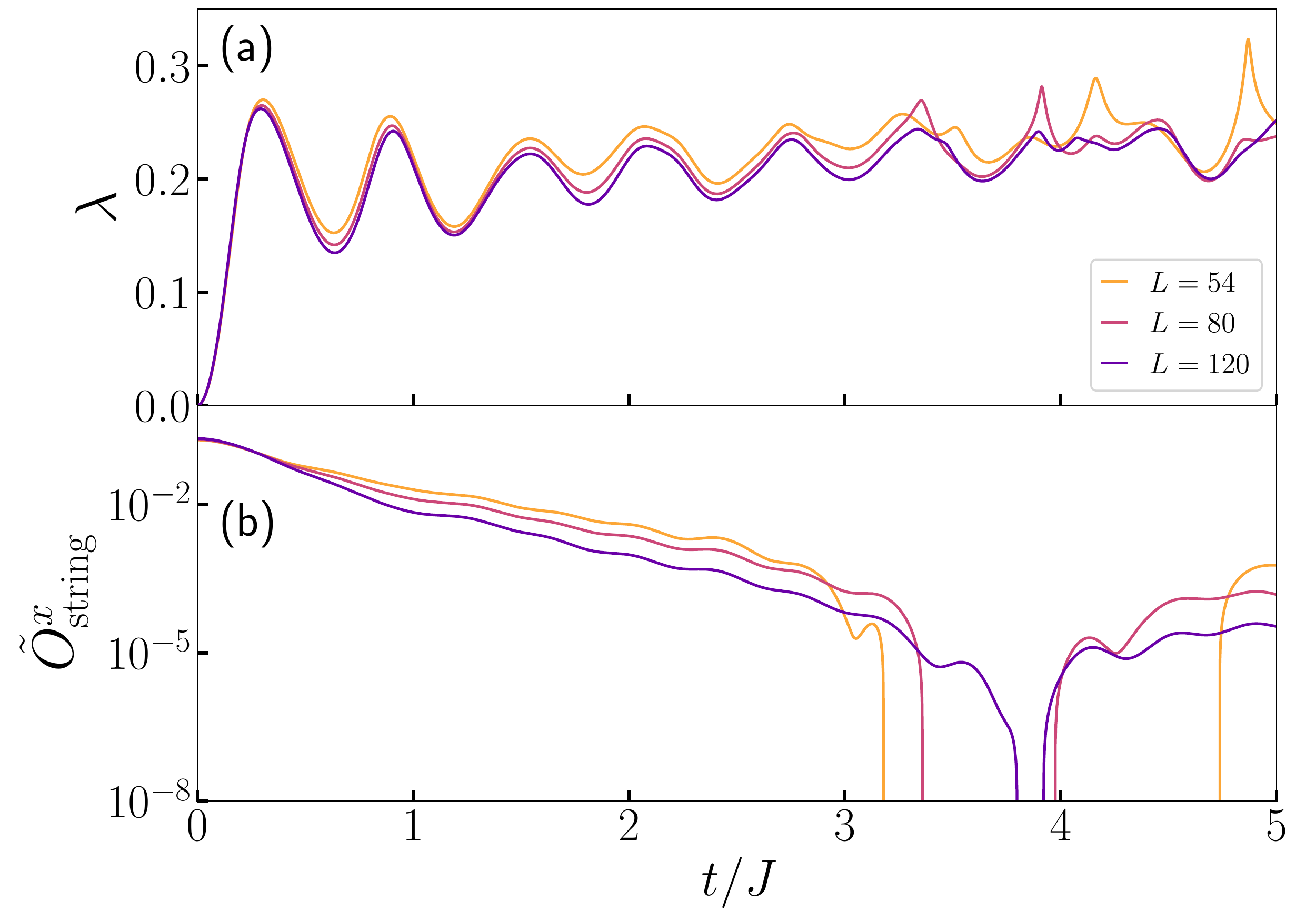}
	\caption{Dynamics of (a) the rate function $\lambda$, and (b) the x string correlation function after a Quench from HI ($V_\text{i}=3.25$) to MI ($V_\text{f}=0.0$) for $L=54,80,120$.}
\end{figure}

\section{Phase diagram and dynamics of spin-1 XXZ chains}
\begin{figure}[h]
	\centering
	\includegraphics[width=0.7\textwidth]{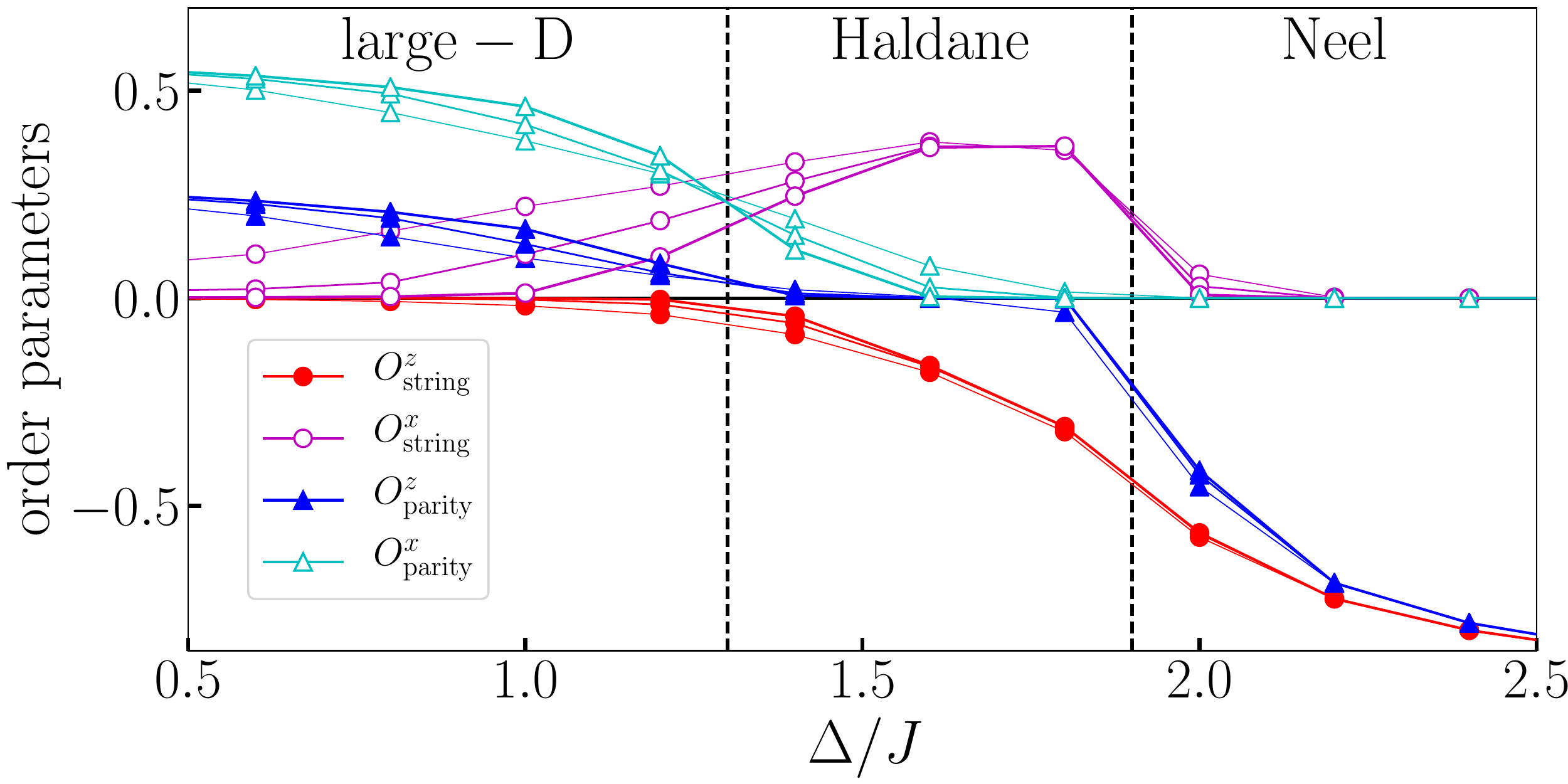}
	\caption{Order parameters of the spin-1 chain as a function of $V$ for $D=1.25$ and system sizes $L=40,80,160$, increasing from thin to thick lines.}
	\label{fig:spinDMRG}
\end{figure}

A spin-$1$ XXZ chain with Hamiltonian
\begin{equation}
	H_\text{XXZ} = J \sum_{i=1}^{L-1} \left( S_i^x S_{i+1}^x + S_i^y S_{i+1}^y \right)
	+ D \sum_{i=1}^{L-1} (S_i^z)^2 + \Delta \sum_{i=1}^L S_i^z S_{i+1}^z
	\label{eq:XXZ}
\end{equation}
can serve as an effective model for the extended Bose-Hubbard model, when the on-site interaction $U$ is large. As a benchmark for how well the effective description works, we show a phase diagram and the corresponding order parameters in Fig.~\ref{fig:spinDMRG}. String and parity order parameters for the spin-1 chain are given by
\begin{equation}
	\begin{split}
		O_\text{string}^\alpha &= \lim_{|i-j|\to\infty} S_i^\alpha \prod_{i < k < j} e^{i\pi S_k^\alpha} S_j^\alpha ,\\
		O_\text{parity}^\alpha &= \lim_{|i-j|\to\infty} \prod_{i < k < j} e^{i\pi S_k^\alpha},
	\end{split}
\end{equation}
where $\alpha=x,z$. There is a rough correspondence $D \to 1/4 \,U$ and $\Delta \to 1/2 \,V$, when $J=1$ in both models, and if one replaces $a_i^\dagger \to S_i^-$. Thus, from $U=5$ in the bosonic model, we obtain $D=1.25$. Along this line, one moves from the large-D phase (corresponding to MI) for small $\Delta$ through the Haldane phase into the Neel phase (corresponding to CDW) for large $\Delta$. The phase transition points are at $\Delta^\text{eq}_{c1} = 1.3 \pm 0.1$ and $\Delta^\text{eq}_{c1} = 1.9 \pm 0.1$. Qualitatively, we have a very good agreement between the models, while numerical values obviously differ.

Finally let us briefly touch on the signs before $J$ in the two models (Here, we always take $J$ to be positive). We take the sign before $J$ in the EBHM negative, while in Eq.~\eqref{eq:XXZ} it is positive as in Ref.~\cite{hagymasiDynamicalTopologicalQuantum2019}. However, the difference is not essential. We have checked the case of an XXZ chain with the negative sign and found that---both for static and dynamical properties---the only difference is a staggered sign $(-1)^{|i-j|}$ that has to be included into the string and parity correlation functions. Therefore, the results are basically unaffected by the sign. A related discussion can be found in Ref.~\cite{pollmannEntanglementSpectrumTopological2010}, distinguishing the Haldane phases with opposite signs as being protected by different symmetries. While for the positive sign the symmetry is a regular lattice inversion symmetry, for the negative sign this symmetry needs to be slightly modified.

\begin{figure}[h]
	\centering
	
	\begin{minipage}{0.45\textwidth}
		\centering
		\includegraphics[width=\textwidth]{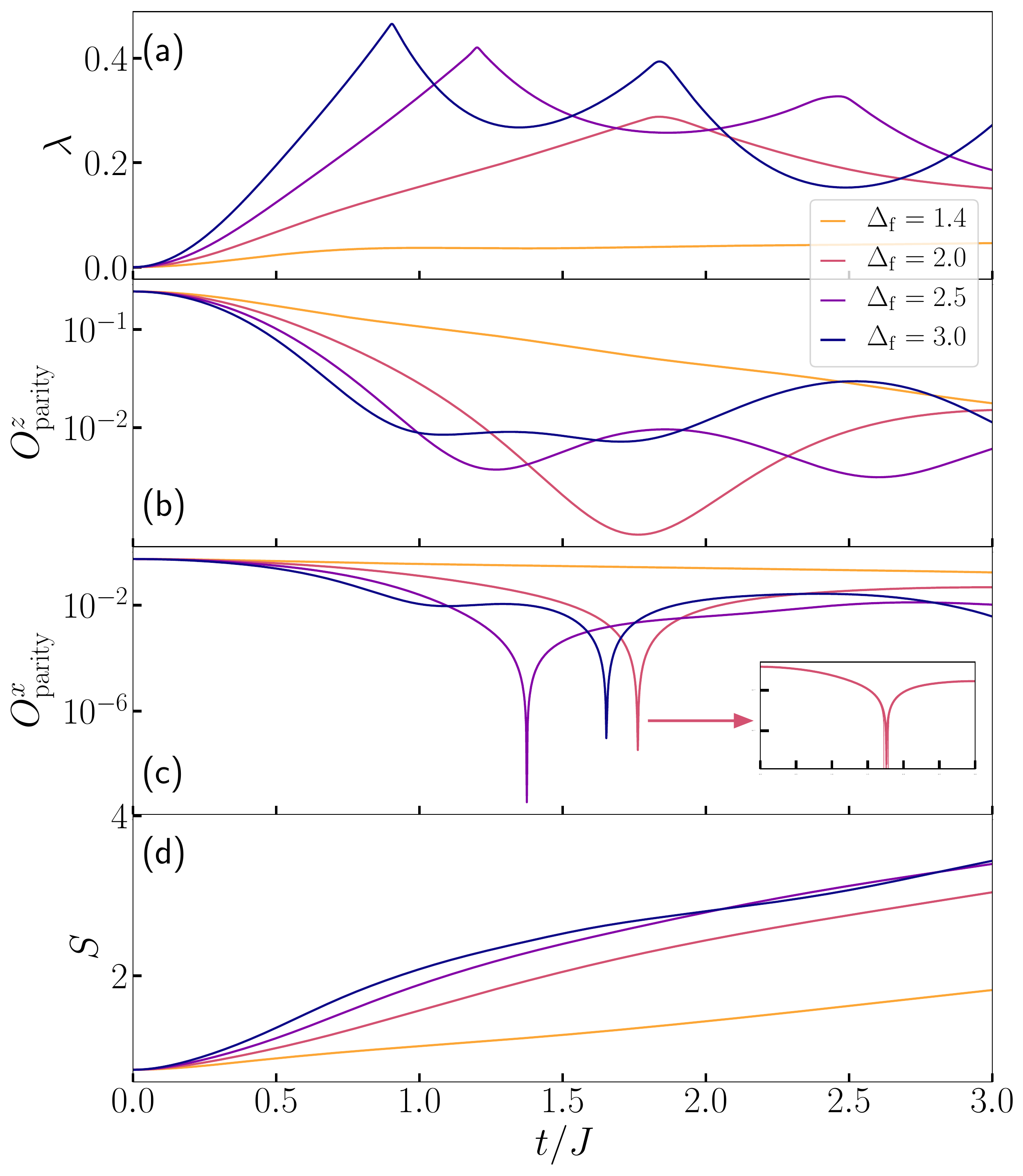}
		\caption{Time evolution of (a) the rate function, (b) and (c) the parity order parameters, and (d) the entanglement entropy in a spin-$1$ XXZ chain starting from an initial state in the large-D phase ($\Delta_i = 0.5$). The inset of panel (c) shows the system size dependence of $O_\text{parity}^x$ for $\Delta_\text{f}=2.0$. 
		}
		\label{fig:spinMI}
	\end{minipage}
	\hfill
	\begin{minipage}{0.45\textwidth}
		\centering
		\includegraphics[width=\textwidth]{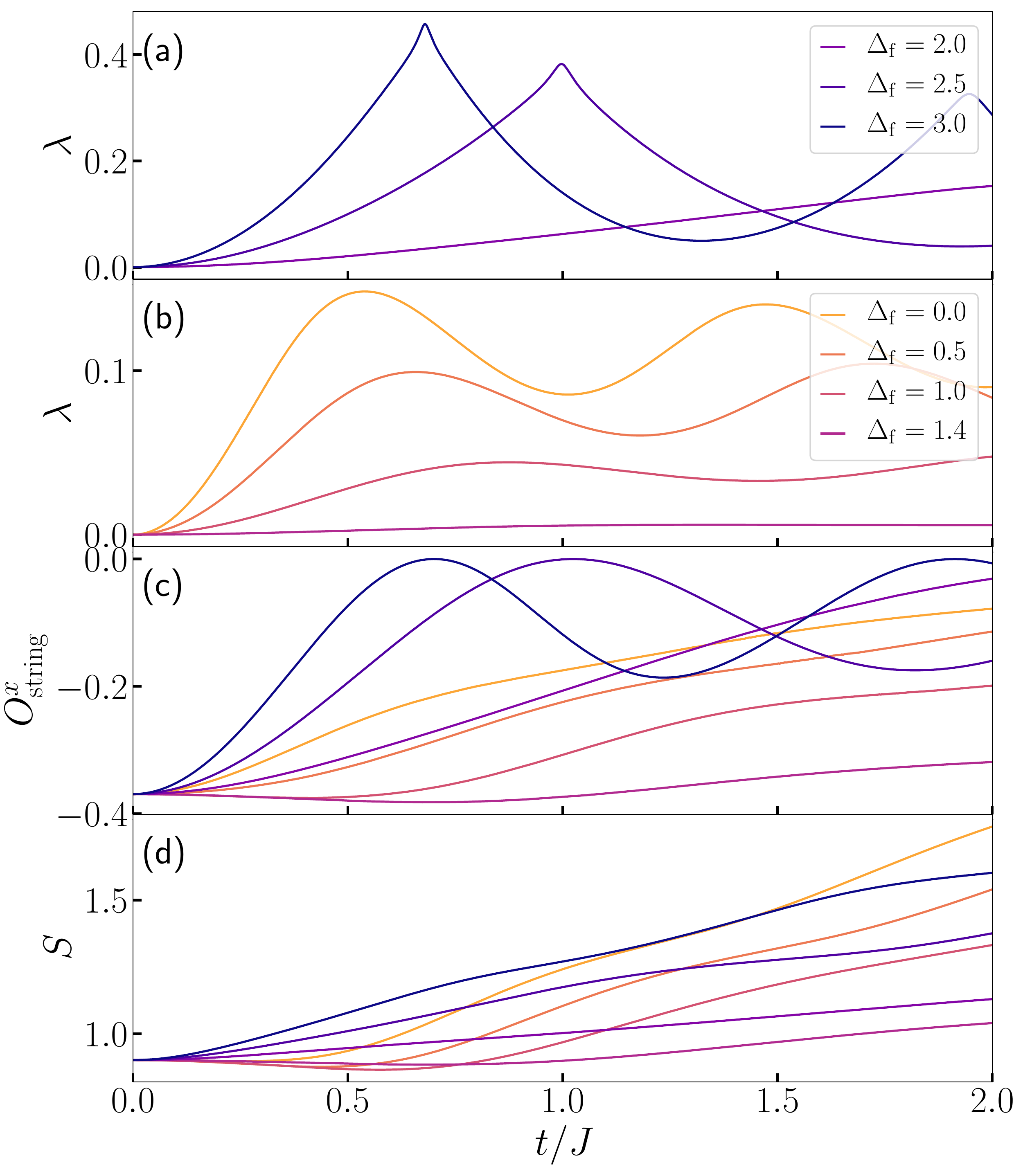}
		\caption{Time evolution of (a) the rate function, (b) and (c) the string order parameters, and (d) the entaglement entropy in a spin-$1$ XXZ chain starting from an initial state in the Haldane phase ($\Delta_i = 1.6$). \vspace{2\baselineskip} }
		\label{fig:spinHI}
	\end{minipage}
\end{figure}

Let us also compare the dynamics of the EBHM to the spin-1 model. We fix $J=1$ and $D=1.25$. As an analog to the initial MI ground state we choose $\Delta_\text{i}=0.5$ in the large-D phase (see Fig.~\ref{fig:spinMI}), and for the initial HI ground state we choose $\Delta_\text{i}=1.6$ (see Fig.~\ref{fig:spinHI}). These initial values are based on Fig.~\ref{fig:spinDMRG}.

After starting in the large-D phase, we find the same kind of DQPTs as in the EBHM when the quench is strong. The threshold for DQPTs, $1.4 < \Delta_\text{th} < 2.0 $, does not coincide with the phase transition points $\Delta^\text{eq}_{c1}$ and $\Delta^\text{eq}_{c2}$ [see Fig.\ref{fig:spinMI}(a)]. More importantly, as shown in Fig.~\ref{fig:spinMI}(c), a sharp minimum of $O_\text{parity}^x$ appears at $\Delta_\text{f} = \Delta_\text{th}$ at the same time as the first kink of the rate function $\lambda$. For $\Delta_\text{f} > \Delta_\text{th}$, e.g. $\Delta_\text{f}=3.0$, this minimum of $O_\text{parity}^x$ occurs at later times than the first kink of $\lambda$. The finite size dependence of $O_\text{parity}^x$ is depicted in the inset of Fig.\ref{fig:spinMI}(c). In contrast to the projected $x$-parity operator of the EBHM in the main text, it converges in the system size.

Lastly, the dynamics starting from a state in the Haldane phase is shown in Fig.~\ref{fig:spinHI}. We find qualitative agreement with the EBHM in the main text. In particular, as can be seen from Fig.~\ref{fig:spinHI}(b) for quenches towards smaller $\Delta$'s, there are no DQPTs on the time scale of the simulation, while the rate function does show several smooth maxima. This confirms that the present initial state, which has a lower symmetry than the ground state of the Heisenberg model ($J=1$, $D=0$, $V=1$), has a significant influence on the timescales of DQPTs.

\section{Spectrum}
\begin{figure}[h]
	\centering
	\includegraphics[width=\linewidth]{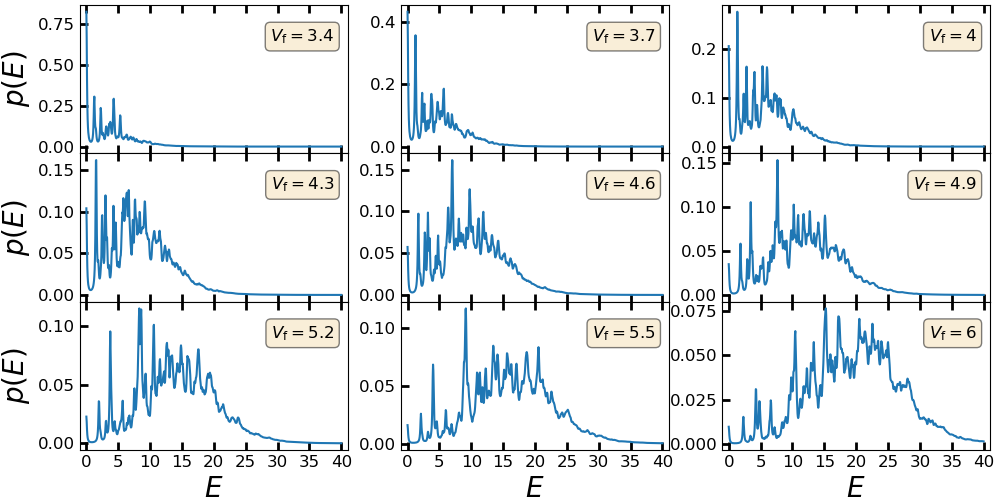}
	\caption{Overlaps of the ground state for $V_\text{i}=1.0$ with the eigenstates for various $V_\text{f}$ as a function of the energy eigenvalues. The system size is $L=10$ and the peaks are represented as Lorentzians of width $\eta=0.1$. Exact diagonalization and open boundary conditions are used.}
	\label{fig:spectrum10MI}
\end{figure}

\begin{figure}[h]
	\centering
	\includegraphics[width=\linewidth]{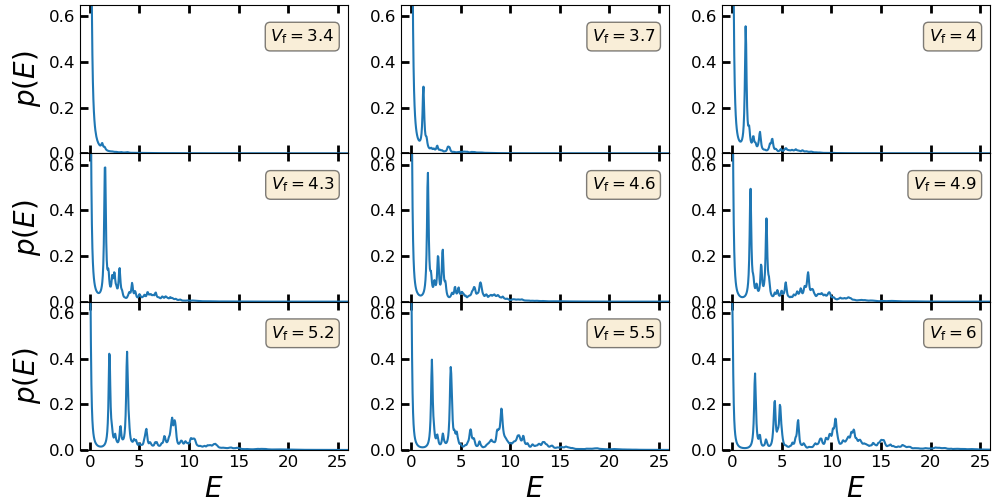}
	\caption{Overlaps of the ground state for $V_\text{i}=3.25$ with the eigenstates for various $V_\text{f}$ as a function of the energy eigenvalues. The system size is $L=10$ and the peaks are represented as Lorentzians of width $\eta=0.1$. Exact diagonalization and open boundary conditions are used.}
	\label{fig:spectrum10HI}
\end{figure}

In order to show how the energy is brought into the system by a quench, we have a closer look at the energy spectrum of the initial state $\ket{\psi_\text{i}}$ with respect to the post-quench Hamiltonian $H_\text{f}$, i.e.,
\begin{equation}
	\ket{\psi_\text{i}} = \sum_{j} c_j \ket{E_j^\text{f}},
\end{equation}
where $H_\text{f} \ket{E_j^\text{f}} = E_j^\text{f} \ket{E_j^\text{f}}$. In the continuum limit, the overlap of the time evolved and the initial wavefunction can be written as
\begin{equation}
	\braket{\psi(t)|\psi_\text{i}} = \int d\varepsilon \, p(\varepsilon) e^{-i\varepsilon t}.
\end{equation}
The distribution $p(\varepsilon)$ is given by
\begin{equation}
	p(\varepsilon) = \sum_j |c_j|^2 \delta(\varepsilon-E_j^\text{f}).
\end{equation}
As this is difficult to obtain in the continuum limit, we have calculated the overlap amplitudes $c_j$ by full diagonalization for $L=10$, and approximated $\delta(\varepsilon-E_j^\text{f})$ by a Lorentzian of width $\eta=0.1$. Figs.~\ref{fig:spectrum10MI} and~\ref{fig:spectrum10HI} show these spectra for $V_\text{i}=1.0$ and $V_\text{i}=3.25$, respectively, for various $V_\text{f}$ with the open boundary condition. As $V_\text{f}$ increases, the means gradually shift to higher values and the distributions become wider. For $V_f =6$, we see a clear signature of a gap formation in the spectrum. We checked that the same happens for the periodic boundary condition, where, due to the momentum conservation, the number of eigenstates with a finite overlap with $\ket{\psi_\text{i}}$ is roughly decreased by $1/L$, and the peaks are more sparsely distributed.

A heuristic model that neglects the intricacies of the actual spectra is given by a box with mean $\overline{E}$ and width $\Delta E$,
\begin{equation}
	p_{\overline{E},\Delta E}(\varepsilon) = \frac{1}{\Delta E} \Theta\left( \varepsilon - \left[\overline{E} - \frac{\Delta E}{2}\right] \right) \Theta\left( \left[\overline{E} + \frac{\Delta E}{2}\right] - \varepsilon \right).
\end{equation}
The corresponding Fourier transform is the Loschmidt-amplitude,
\begin{equation}
	\braket{\psi(t)| \psi(0)} = \int_{-\infty}^{\infty} p_{\overline{E},\Delta E}(\varepsilon) e^{-i\varepsilon t} dt = \int_{\overline{E} - \Delta E / 2}^{\overline{E} + \Delta E / 2} \frac{1}{\Delta E} e^{-i\varepsilon t} dt \sim \frac{\sin(\Delta E t)}{\Delta E t}.
\end{equation}
One can see that the times of its zeros scale as $t^* \sim (\Delta E)^{-1}$. Thus we have a most rudimentary example of how the width of $p(\varepsilon)$ controls the times of DQPTs in a power law fashion. As the width of the more complicated spectrum of the EBHM increases for stronger quenches (see Figs.~\ref{fig:spectrum10MI} and~\ref{fig:spectrum10HI}), and the critical times $t^*$ of the EBHM follow approximate power laws as a function of the quench parameter (see Fig.~4 of the main text), the spectrum of the EBHM likely controls the times of DQPTs in a similar way as $p_{\overline{E},\Delta E}(\varepsilon)$ in the above toy model.

%